\documentclass[12pt]{article} 
\usepackage{epsf,subeqn,cite}

\newcommand{\ba}{\begin{array}}
\newcommand{\ea}{\end{array}}
\newcommand{\bd}{\begin{displaymath}}
\newcommand{\ed}{\end{displaymath}}
\newcommand{\be}{\begin{equation}}
\newcommand{\ee}{\end{equation}}
\newcommand{\bea}{\begin{eqnarray}}
\newcommand{\eea}{\end{eqnarray}}

\def\be{\begin{equation}}
\def\ee{\end{equation}}
\def\barr{\begin{array}}
\def\earr{\end{array}}
\def\dis{\displaystyle}
\def\ra{\rightarrow}

\def\RedViolet{} 
\def\Black{}
\def\Blue{}

\def\RawSienna{}

\begin{document}
\setcounter{page}{0}
\thispagestyle{empty}
\setcounter{footnote}{0}
\renewcommand{\thefootnote}{\fnsymbol{footnote}}
\begin{flushright}
{\large MRI-PHY/P981063\\  \tt hep-ph/9810339}
\end{flushright}

\begin{center}
{\Large\bf \Blue 
Semileptonic charm decay as a test for the spectator model in the 
$B_c$-meson}\\[20mm]
\Black
{Debajyoti Choudhury}\footnote{Electronic address: debchou@mri.ernet.in}, 
{Anirban Kundu}\footnote{Electronic address: akundu@mri.ernet.in}
and {Biswarup Mukhopadhyaya}\footnote{Electronic address: 
biswarup@mri.ernet.in}\\[10mm]
{\em Mehta Research Institute,\\
Chhatnag Road, Jhusi, Allahabad - 211 019, India}
\end{center}
\RawSienna
\begin{abstract}
The $b$ and the $c$ quark compete with each other in decays of the
$B_c$-meson. We emphasize the need of obtaining
reliable signals of $c \ra s$ in order to assess the merits of
the spectator approximation in calculating the relative strengths of the two
types of decays. This, we argue, can be done  by considering the decay 
$B_c \ra B_s (B_s^*) l \overline{\nu_l}$, followed by
semileptonic decays of the $B_s$. We suggest looking for like-sign dileptons 
together with a $D_s$ at the BTeV or LHC-B experiments, and show that
such signals can be made background-free by suitable event selection. 
\end{abstract}

\vskip 1 true cm
\Black
\clearpage
\setcounter{page}{1}
\pagestyle{plain}
\setcounter{footnote}{0}
\renewcommand{\thefootnote}{\arabic{footnote}}

The recent discovery of the $B_c$-meson, consisting of a $\overline{b}$ and
a $c$ quark, opens up  rather interesting possibilities whose test may
confront one with some ill-understood aspects of heavy flavour physics
\cite{review_bc}.
So far, the Collider Detector at Fermilab (CDF) have confirmed observations 
of the $B_c$ \cite{cdf_bc}. 
In addition, the ALEPH, OPAL and DELPHI collaborations at the 
Large Electron Positron (LEP) collider at CERN have reported candidates for a 
similar designation \cite{aleph_bc, opal_bc, delphi_bc}. 
The decay channel utilised in all these searches
is one where the $\overline{b}$-quark decays first into a $\overline{c}$,
producing a $J/\psi$, either semi-leptonically or with one or more pions.
This is because  the $J/\psi$ is easily identified through the decay into
two leptons, and the additional leptonic and pionic tracks passing close to 
the decay vertex provide a viable reconstruction of the $B_c$
\cite{atlas}. The best fit 
for the $B_c$ mass obtained from this channel also shows a fair agreement 
with that predicted by potential model calculations \cite{eichten}.

However, even a naive estimate of the lifetimes of the $b$ and the $c$ 
quark shows that, in the $B_c$ meson, the charm decay may take precedence 
over the bottom decay. This is
due to the fact that, unlike $b\ra c$, the  decay $c\ra s$  is not 
suppressed by any off-diagonal 
element of the Cabibbo-Kobayashi-Maskawa (CKM) matrix.
 Such a suppression tends to  offset the effect of the mass of 
the $b$-quark. The annihilation channels of $B_c$ are also important
\cite{ann}.  Both these types of decays have 
already been studied in different theoretical approaches. 
To be more specific, the semileptonic channels $B_c \ra J/\psi 
l \overline{\nu_l}$ and $B_c  \ra B_{s}^{(*)} l \overline{\nu_l}$
have been studied using the spectator approximation and different
types of form-factors, {\em e.g.}, BSW \cite{bsw_bc}, ISGW \cite{isgw},
ACCMM \cite{accmm} etc.   No definite conclusion 
can yet be drawn, however,  regarding the
preferability of a non-relativistic potential \cite{beneke} 
to a relativistic one, when a  $c$ decays into an $s$,
while both of them interact with the b-quark all along \cite{chang}.
It has also been pointed out that the massive 
spectator for the latter mode significantly affects the phase space 
that would have been otherwise 
available with a light spectator \cite{lusignoli}. 
Summing everything up, one can say that the decay widths for the 
two semileptonic modes might as well be  just comparable.

The above remarks show that there is still considerable scope for 
improving our understanding of the $B_c$-system where two heavy quarks 
compete with  each other in the decay processes and at the same time influnece 
the transverse motions of each other.  In order to clearly establish 
the validity 
of a given theoretical picture, it is important at this stage 
to experimentally identify signatures of
decays induced by $c \ra s$.
In this note, we suggest ways of extracting  
the corresponding final states 
in the semileptonic decay  
\be 
\Blue
 B_c  \longrightarrow B_{s}^{(*)} l \overline{\nu_l} \ . 
  \label{signal}
\Black
\ee
One of the reasons for concentrating on semileptonic decays is to ensure that 
our basic question 
does not get swamped by additional theoretical uncertainties concerning, for
example, the legitimacy of factorisation. Thus, our numbers will depend
only on the parametrization of the semileptonic decays using spectator
approximation, and any observed deviation from these numbers will mean 
significant non-spectator physics contribution.

As will be clear from the discussion below, it is 
profitable for us to consider final states which are again produced through 
semileptonic decays of the $B_s$.   
A realistic testing ground for such decays will be provided by 
experiments dedicated to 
B-studies at hadronic colliders, typical examples being the BTeV and LHC-B 
experiments \cite{stone}.
Depending on the integrated luminosity achieved ultimately, these machines
can be expected to produce $10^{11} - 10^{12}$ $b\overline{b}$-pairs per year. 
Thus, with a fragmentation ratio of $3.8\times 10^{-4}$ ($5.4\times
10^{-4}$) for the $B_c$ ($B_c^*$) meson\footnote{$B_c^*$ decays almost
entirely to $B_c$,  so these two numbers just add up to give the total
fragmentation ratio.} \cite{braaten},  
approximately $10^{8}$ -- $10^{9}$ $B_c$s can be  available 
according to a rather conservative estimate. One may expect a 
higher yield  if one includes the possibility
of having a $B_c$ on one side and $B$-meson of a different type on the other
\cite{mangano}.

The decay
$B_c  \ra B_{s}^{(*)} l \overline{\nu_l}$ is followed by the 
$B_s$ decaying  into one of its allowed channels. (The $B^*_s$  decays almost 
entirely into a $B_s$.)  
These include final states consisting of $D_s \pi$ ,   
$D_s 3 \pi$ as well as the semileptonic channel $D_{s}^{(*)} l \nu_l$.  
In this letter. we propose to utilise the semileptonic channel alone. 
Despite suffering from a small branching fraction, this has 
an unique advantage. Because of the 
high rate of $B_{s}-\overline{B_s}$ oscillation, both set of 
final states $D_{s}^{+(*)} l^{-} \overline{\nu_l}$ and 
$ D_{s}^{-(*)} l^{+} \nu_l$ are equally probable. 
Thus in approximately half the cases one expects
like-sign dileptons together with a $D_{s}$ (including one that may have a
$D^*_{s}$ as its source). This kind of a signal is relatively background-free
compared to both the pionic final states and those comprising unlike-sign
dileptons. Sources that can fake such signals are 
limited to those which have two-level
neutral particles at the last step of a cascade. The most prominent example
of such  backgrounds is the decay chain
$B^0 \ra D_{s}^{(*)} D^{-} \ra D_{s} K^{0} 
l^{-} \nu$, following which the $K^0$ can give rise to leptons of either
sign. We shall show that they can  be rather easily removed by kinematic cuts.

Calculation of the rates for the exclusive semileptonic decays is 
straightforward in a form-factor approach.
The hadronic matrix elements relevant for the decay of a 
generic pseudoscalar meson $P_0$ may be parametrized 
as
\be
\barr{rcl}
  \langle P(k) | \bar q_1 \gamma_\mu (1 - \gamma_5) q_2
                                          | P_0(p) \rangle & = & 
               f_+ \;  (p + k)_\mu  + f_- \;  q_\mu
         \\[2ex]
 \langle V(k,\epsilon) | 
            \bar q_1 \gamma_\mu (1 - \gamma_5) q_2
                                          | P_0(p) \rangle 
         & = & \dis 
               \frac{2 V_{P_0 V} \; }{m_0 + m_V} 
                    \epsilon_{\mu \nu \alpha \beta} \epsilon^{\ast \nu} 
                 p^\alpha k^\beta 
         \\[1.5ex]
         & + & \dis i
                \Bigg[ (m_0 + m_V) \; A^{(1)}_{P_0 V} \;  \epsilon^{\ast}_\mu 
         \\[1.5ex]
         &  & \dis \hspace*{1em}
                    - \epsilon^{\ast} \cdot p 
                           \left\{ \frac{A^{(2)}_{P_0 V} \; }{m_0 + m_V} 
                              \; (k + p)_\mu 
                    + 2 \frac{m_V}{q^2}
                    \left[
                     A^{(3)}_{P_0 V} - A^{(0)}_{P_0 V} 
                      \right] q_\mu 
                           \right\}
                 \Bigg]  
\earr
         \label{formfac}
\ee
In eq.(\ref{formfac}), $q \equiv p - k$ and  $P$ and $V$
correspond to generic pseudoscalar and vector mesons respectively.
The value of the form-factors $f_\pm$, $V$ and $A^{(i)}$ 
and their $q^2$--dependence can be calculated within a specific 
model. For the rest of our analysis, we shall use the results obtained 
within the Bauer-Stech-Wirbel (BSW) formalism \cite{bsw}
which we compile in Table \ref{tab:formfactors}.
We have checked that our results are insensitive
to, say, a 20\% variation of the form-factors, including that arising from
the uncertainty in the parameter $\omega$ in the BSW framework. 
%
\begin{table}[htb]
\begin{center}
\begin{tabular}{|l|c|c|c|c|c|c|c|c|}
\hline
Mode& $m_b$& $m_c$& $m_s$& $\omega$& $h_0$& $h_V$& $h_{A_1}$& $h_{A_2}$\\
\hline
$B_c\ra B_s$& 4.9& 1.7 & 0.55& 0.4& 0.403& --& --& --\\
& & & & 0.5& 0.588& --& --& --\\
& 5.0& 1.5& 0.5& 0.4& 0.432& --& --& --\\
& & & & 0.5& 0.617& --& --& --\\
$B_c\ra B_s^*$& 4.9& 1.7 & 0.55& 0.4& --& 3.367& 0.487& 1.155\\
& & & & 0.5& --& 4.794& 0.693& 1.697\\
& 5.0& 1.5& 0.5& 0.4& --& 3.527& 0.521& 1.246\\
& & & & 0.5& --& 4.905& 0.725& 1.765\\
\hline
\end{tabular}
\end{center}
\caption{\em Form-factors relevant to the decays of
eq.(\protect\ref{signal}) as calculated within the
BSW formalism~\protect\cite{bsw}. The values are shown for
$q^2=0$ only. The relevant $c\bar s$ poles are: $2.60$ GeV
$(f^+)$, $2.11$ GeV $(V_{P_0V})$, $2.53$ GeV $(A^{(1,2)}_{P_0V})$.
The other formfactors do not contribute in the limit of vanishing
lepton masses. All masses and $\omega$ are in GeVs. All other 
form-factors are obtained from Ref.\protect\cite{bsw}.
}                                                              
    \label{tab:formfactors}
\end{table}
%

In the remainder of our analysis we shall assume that the 
pions and/or photons resulting from the decay of an excited 
state(s) to the corresponding ground state will remain untagged.
Since the signal consists of a $D_s$ 
together with a pair of like-sign dileptons, 
the relevant backgrounds emanate from processes that 
include these particles in the final state apart from 
other undetected ones.  The most important 
processes\footnote{There also exist additional channels with 
          intermediate $D_s^\ast$ and/or $D^{\ast -}$ states 
          cascading into the respective ground states. Since the 
          phase space distributions are quite akin to those above, 
          these channels cannot be distinguished unless the 
          accompanying photon(s) are detectable. We shall assume 
          that this is not the case and add on such contributions 
          to those of eqs.(\protect\ref{bkgd_1}) and (\protect\ref{bkgd_2}).}
in this context are
\subequations
\be
\RawSienna
B^0_d \longrightarrow D_{s} D^{-} \longrightarrow 
                      D_{s} K^{0} l^{-} \nu 
           \ ,
       \label{bkgd_1}
\Black
\ee
and
\be
\RawSienna
B^0_d \longrightarrow D_{s} D^{-} \longrightarrow 
                      D_{s} K^{\ast 0} l^{-} \nu 
                      \longrightarrow 
                      D_{s} K^{0} \pi l^{-} \nu  
           \ .
       \label{bkgd_2}
\Black
\ee 
\endsubequations
The $K^0$ has a 50\% probability each of being in the 
$K_L$ or the $K_S$ state. With the $K_S$ having a negligibly small 
semileptonic branching ratio, it is the $K_L$ which can decay 
(BR $\simeq 65 \%$) into $\pi l \nu_l$ ($l = e, \mu$),  $l^+$ and $l^-$
being produced with equal probability.

As $m(D_s) - m(K^{\ast 0})$ is considerably smaller than 
$m(D_s) - m(K^0)$, 
lepton from the $D_s$ decay will have significantly different
energy spectra in the two cases. Similarly, the leptons 
in the signal events will have an entirely different profile. 
We aim next to quantify this difference in a frame-independent 
manner. 

%
\begin{figure}[htb]
\vspace*{-0.0cm}
\begin{center}
\epsfxsize=9.3cm\epsfysize=7.3cm
\epsfbox{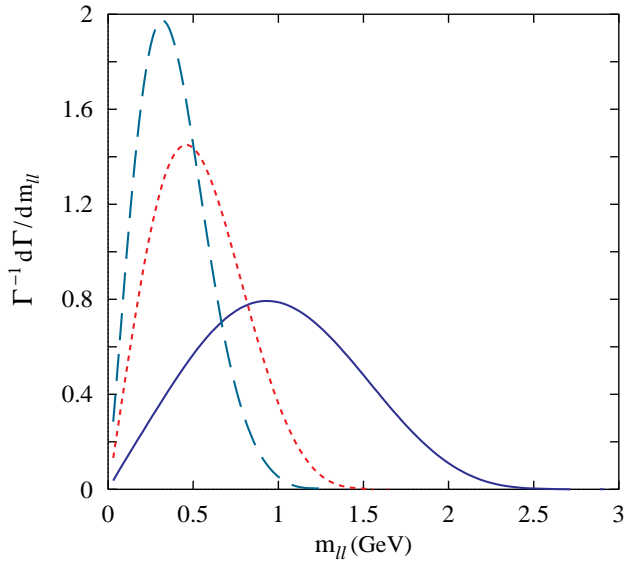} 
\vspace*{-0.3cm}
\end{center}
\caption[fig:fig1]{\em Distribution in the invariant mass of the
lepton pair. The solid line corresponds to the signal, while the
short- and long-dashed lines refer to backgrounds from the decay 
chains (\protect\ref{bkgd_1}) and (\protect\ref{bkgd_2}) respectively.
} \label{fig:m_ll}
\end{figure}
%
The first kinematic distribution on which we focus our attention is 
$\Gamma^{-1} \; {\rm d}\Gamma / {\rm d} m_{\ell \ell}$ where 
$m_{\ell \ell}$ is the invariant mass 
of the dilepton system. Note that this is a Lorentz invariant distribution and
is unaffected by the extent to which the parent $B_c$s are boosted. 
For the signal, this distribution shows a peak in the neighbourhood 
of 1 GeV (see Fig.~\ref{fig:m_ll}),
and has little dependence on the form-factors.  
The figure  clearly shows that the background peaks at a considerably
lower invariant mass compared to the signals. This is because each of the two 
leptons in the background process is produced at a later stage of the
cascade, thereby suffering a degradation that lowers the invariant 
mass. 

\begin{figure}[htb]
\vspace*{-0.0cm}
\begin{center}
\epsfxsize=9.3cm\epsfysize=7.3cm
\epsfbox{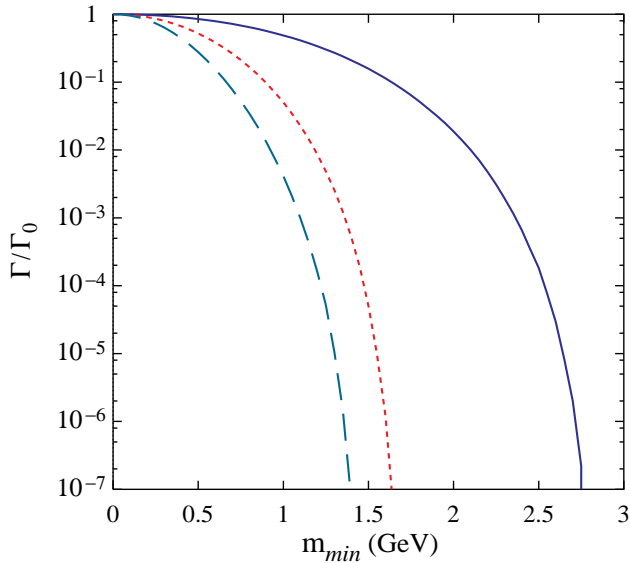}  
\vspace*{-0.3cm}
\end{center}
\caption[fig:fig2]{\em The fraction of the signal and background 
retained on imposition of a lower bound on the invariant mass of the
lepton pair. The solid line corresponds to the signal, while the
short- and long-dashed lines refer to backgrounds from the decay 
chains (\protect\ref{bkgd_1}) and (\protect\ref{bkgd_2}) respectively.
} \label{fig:inv_cut}
\end{figure}

It must be borne in mind though, that Fig.~\ref{fig:m_ll} exhibits
only the normalized distributions. However, at a hadronic machine, 
the  production rate for $B^0_d$ is nearly 3 orders 
of magnitude higher than that for $B_c$. Again, while the 
branching fractions for the backgrounds of 
eqs. (\ref{bkgd_1}) and (\ref{bkgd_2}) are 
$1.1 \times 10^{-3}$ and $4.7 \times 10^{-4}$ respectively \cite{pdg}, 
that for the signal chain is 
approximately\footnote{Note that this includes the factor of half 
        due to the fact that we consider only the like-sign 
        dilepton case.}
$0.08 \: {\cal B}$ where 
\be 
\Blue
    {\cal B} = Br (B_c \longrightarrow B_s \ell \nu) + 
        Br (B_c \longrightarrow B_s^\ast \ell \nu)  \ .
   \label{brfrac}
\Black
\ee
Using the theoretical prediction of ${\cal B} \sim 0.2$ \cite{review_bc},
it is easy to see that 
the background can, {\em a priori}, be much larger than the signal.
However, as Fig.~\ref{fig:inv_cut} shows, 
one could still separate the signal from the background by
imposing a cut\footnote{The invariant mass for the 
                 signal falls far short of the $J/\psi$ and hence 
                 such backgrounds can easily be eliminated without loosing 
                 any of the signal.}
on the invariant mass for the dilepton pair:
\[
\RawSienna
      m_{\ell \ell} > m_{min} \ .
\Black
\]
%
%
Clearly, with a strong cut on $m_{\ell \ell}$ most of the background 
can be eliminated while losing a relatively smaller fraction of the 
signal. For example, with a $m_{min}$ of  $0.8$, $1.2$ and $1.6$ GeV, 
one retains 
\RedViolet
64.5\%,  33.8\% 
\Black
and 
\RedViolet
11.4\% 
\Black
of the 
signal respectively. For the same set of cuts, 
the background (\ref{bkgd_1}) gets reduced to 
\RawSienna 
16.7\%, 0.86\% 
\Black
and 
\RawSienna 
0.001\% 
\Black
respectively, while  background (\ref{bkgd_2}) falls to 
\RawSienna 
3.4\%, 0.01\% 
\Black
and 
\RawSienna 
0\% 
\Black
of its unrestricted value. 

Although a relatively large value of $m_{min}$ would 
eliminate all of the background, 
one needs to ensure that the events retained are sufficiently
numerous for them to be considered a signal. To make a judicious
choice, one thus has to consider the event rates expected. 
The latter is obviously given by the product 
$(0.08 \: {\cal B} N \epsilon) $, 
where $N$ is the 
number  of $B_c$s produced,  and $\epsilon$ is the experimental 
efficiency of detecting the final state. Now, $\epsilon$ is the 
same for both the signal and the background, while the number 
of $B_d$s produced can be 
approximated\footnote{Approximately only 0.1\% of the $b$ quarks produced
           end up in $B_c$s while most of them hadronize into 
           $B_d$ and $B^\pm$  with nearly equal probability.}
to be $500 N$. 

At the BTeV experiment, 
$N~\simeq~10^8$ even with a conservative
estimate. The rate may be slightly higher at the LHC-B. The  identification 
efficiency for each lepton is about 90 \%. The $D_s$ is mostly to be
recognised through the $\phi \pi$ and $K^{*}K$ channels, in which the 
combined branching ratio is 6\%. Each of these will give rise to three tracks. 
Including the trigger efficiency and all kinematic cuts necessary for proper
identification, the average efficiency for these three tracks being faithfully 
 reconstructed into a $D_s$ is approximately 2.5\%. Thus, with the requirement
of proper identification of the other lepton (coming from $B_c$-decay),
the net efficiency of detection of the final states of our concern 
turns out to be $\epsilon \sim 0.001$. 
It should be noted that the parent $B_c$ is 
sufficiently boosted, so that the lepton coming from its decay at the first 
step stays more or less in the same rapidity interval as the $D_s$ decay
tracks and the lepton emanating from $B_s$-decay. 

In Table~\ref{tab:signif}, we present, for a given $N \epsilon$,
the number of signal ($S$) and background ($B$) events as a function 
of $m_{min}$. The significance $S/\sqrt{B}$ (which is proportional to
$\sqrt{N\epsilon}$) is then easy to 
calculate.
%
\begin{table}[h]
\begin{center}
\begin{tabular}{ | c | r | r | r |}
\hline
$m_{min}$  & \multicolumn{1}{|c|}{Signal} & \multicolumn{2}{|c|}{Background} 
          \\[1.1ex]
\cline{3-4}
(GeV)& & \multicolumn{1}{|c|}{(\protect\ref{bkgd_1})} & 
         \multicolumn{1}{|c|}{(\protect\ref{bkgd_2})}
          \\
\hline
0.0   &  \RedViolet 8000 \Black $ {\cal B} $  & 55000 &  23500 \\
0.2   &  \RedViolet 7820 \Black $ {\cal B} $  & 49980 &  19330 \\
0.4   &  \RedViolet 7240 \Black $ {\cal B} $  & 36610 &  10410 \\
0.6   &  \RedViolet 6320 \Black $ {\cal B} $  & 21100 &   3640 \\
0.8   &  \RedViolet 5160 \Black $ {\cal B} $  &  9180 &    813 \\
1.0   &  \RedViolet 3900 \Black $ {\cal B} $  &  2780 &     98 \\
1.2   &  \RedViolet 2700 \Black $ {\cal B} $  &   474 &    3.7 \\
1.4   &  \RedViolet 1770 \Black $ {\cal B} $  &    28 &      0 \\
1.6   &  \RedViolet  910 \Black $ {\cal B} $  &     0 &      0 \\
\hline
\end{tabular}
\end{center}
\caption{\em The number of signal and background events retained 
             as a function of $m_{min}$ for $N \epsilon = 10^5$. 
             We assume that 500 as many $B_d$s are produced as 
             $B_c$s. }
    \label{tab:signif}
\end{table}
%
Since the branching ratios for all other decays in 
both the signal and the background cascades 
are rather well-known, it should be possible  from 
the above estimates to measure ${\cal B}$, the sum of the 
branching  ratios for 
$B_c  \ra B_{s}^{(*)} l \overline{\nu}$  and match it with
the different theoretical predictions. It is interesting to
note that, even without imposing any kinematical cuts, it 
would be possible to have a $5 \sigma$ signal 
as long as ${\cal B} > 0.17$.

Further distinction of the signal from the backgrounds can be achieved by
considering different angular distributions. These distributions in the
laboratory frame require the $B_c$ to be boosted appropriately,
and one can perform a detailed simulation only by knowing the machine 
parameters. On the other hand, some interesting  features can be observed for 
the distributions in the rest frame of the $B_c$. 

%
\begin{figure}[htb]
\vspace*{-1.5cm}
\hspace*{-2.3cm}
\begin{center}
\epsfxsize=6.55cm\epsfysize=6.6cm
\epsfbox{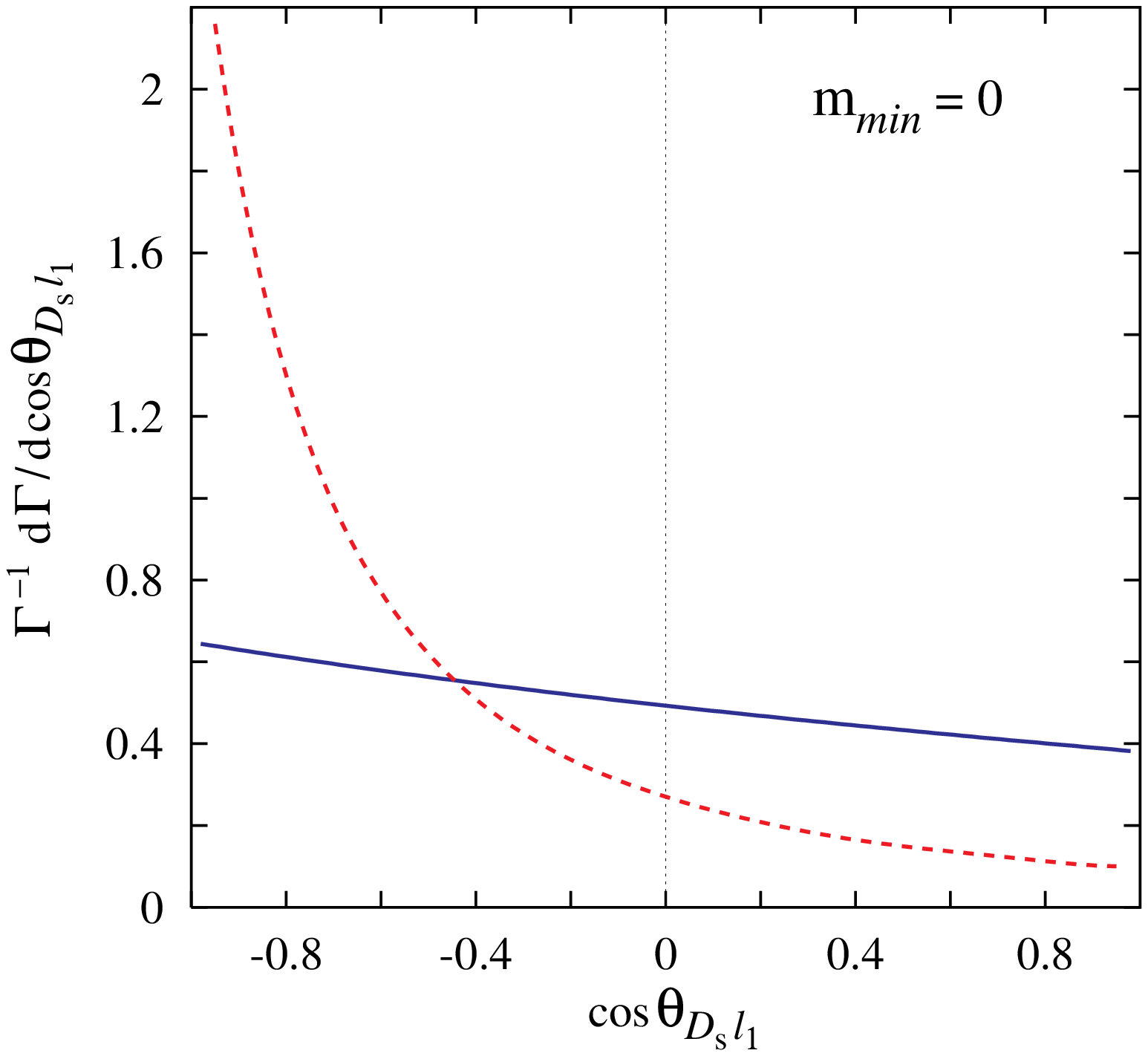}
\hspace*{-1.65cm}
\epsfxsize=6.55cm\epsfysize=6.6cm
\epsfbox{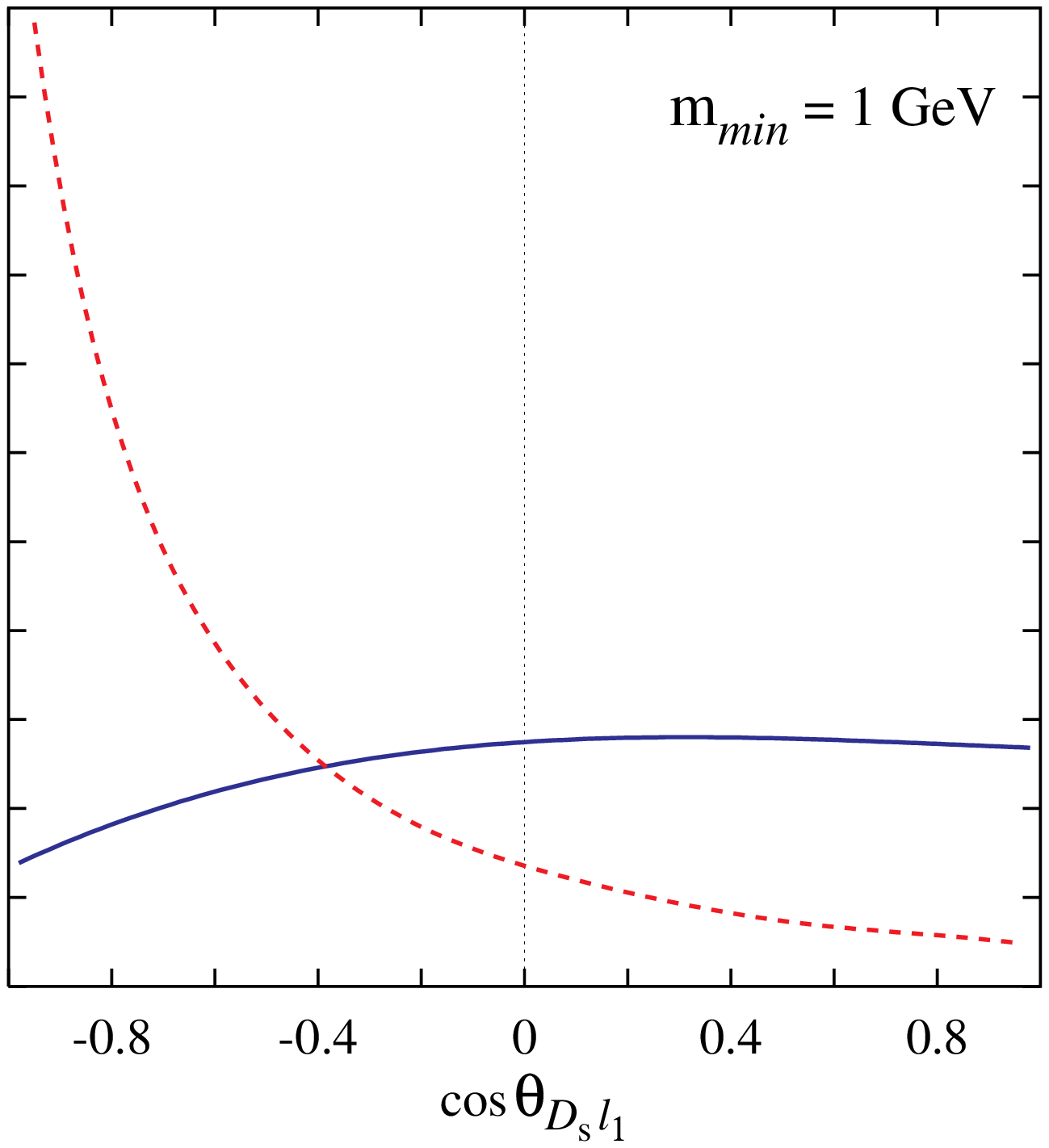}
\hspace*{-1.65cm}
\epsfxsize=6.55cm\epsfysize=6.6cm
\epsfbox{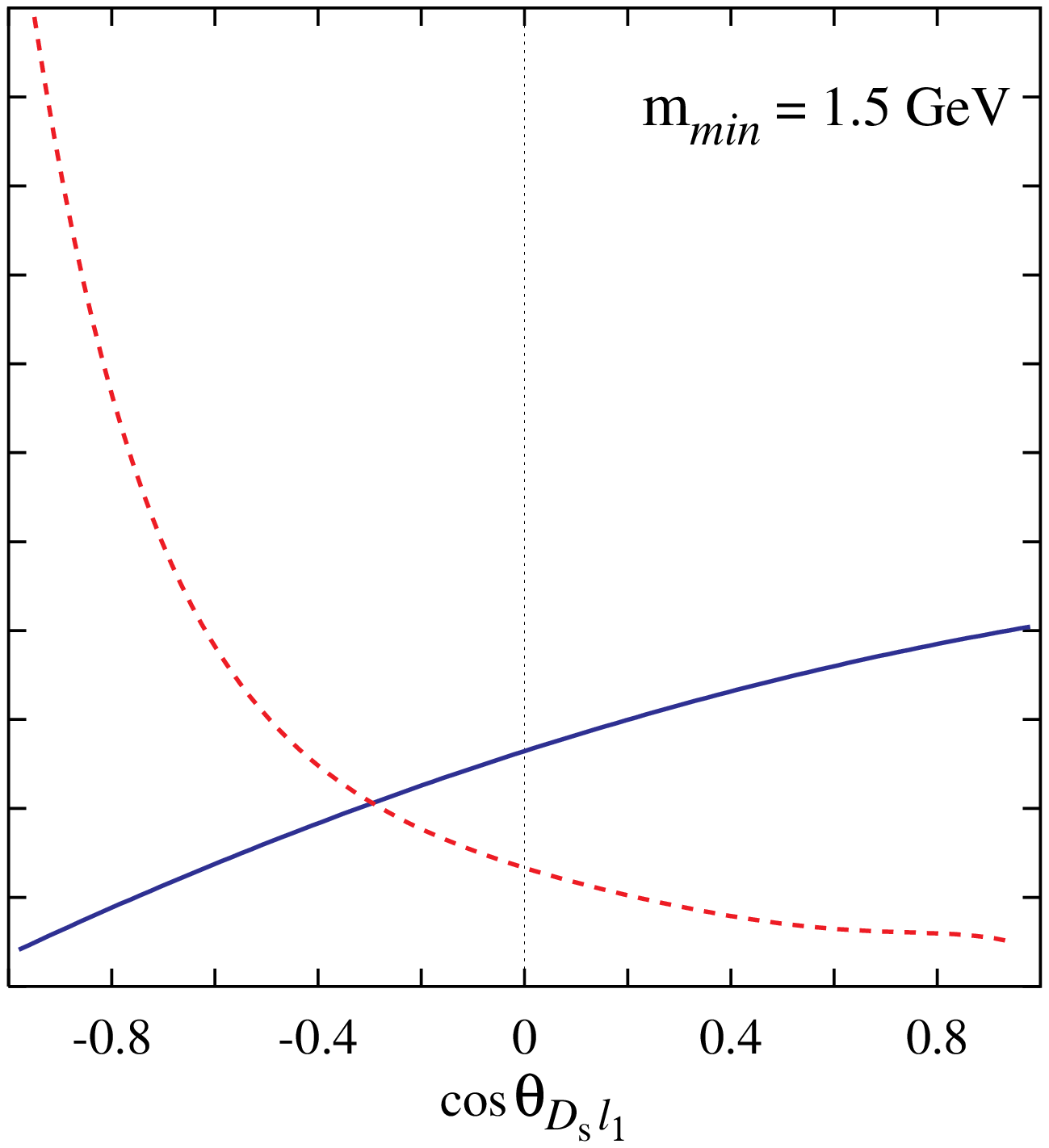}   
\vspace*{-0.6cm}
\end{center}
\caption[fig:fig3]{\em The distribution in the angle between the 
        $D_s$ and the slower of the two leptons for 
        different values of the lower bound on the dilepton 
        invariant mass. The solid line corresponds to the signal, while the
        dashed line refers to the background (the curves for the 
        two backgrounds (\protect\ref{bkgd_1}) and (\protect\ref{bkgd_2}) 
        are virtually indistinguishable).
        } 
\label{fig:angular}
\end{figure}
%

Let us, for example, consider, for the signal process 
$B_c\ra B_s (B_s^*)$,
the distribution in the opening angle between the
reconstructed $D_s$ and the slower of the two leptons (which, in most 
of the cases, is the one coming from the primary decay). 
As shown in figure 3, the distribution
is a slowly falling one, as opposed to the corresponding case for the
backgrounds, where a peaking for large angle is evident. This can be
explained by noting that the background leptons originate from
$B^0_d \ra D_{s}(D^{*}_{s}) D^{-}$ followed by cascade decay
of the $D^{-}$. Since the momentum of the lepton in the rest frame 
of the $D^-$ is smaller than the momentum of the $D^-$ itself, 
the leptons have trajectories close to that of the $D^{-}$. 
The latter, on the other hand,
moves back-to-back with respect to the $D_s$ in the $B_c$ rest frame, 
making it imperataive for the leptons, too,  to have a large opening angle 
against the $D_s$. 

With increasing $m_{min}$, events with more energetic leptons, 
or with large opening angles between them, are filtered out.
For the signal events,  
this results in a slightly enhanced  likelihood of small opening
angle between the slow lepton and the $D_s$. For the background,
on the other hand, an increase is $m_{min}$ does not change 
this particular distribution to a perceptible extent. This 
can be traced to the event topology discussed in the last 
paragraph.

Though the distinctions suggested above are most noticeable in the rest frame
of the $B_c$, they can be utilised if the latter can be reconstructed 
even at a statistical level. If such reconstruction is possible, 
the study of the angular distributions will enable one to filter out the
signals even with a somewhat  relaxed invariant mass cut and thus to
be left with a larger number of events.

In conclusion, the study of $D_s$ plus like-sign dilepton signals can serve
as extremely useful pointers to the the decay $c \ra s$ 
taking precedence over $b \ra c$ in a $B_c$ meson. Using
invariant mass cuts on the dileptons, one can reduce the main backgrounds
to a large extent while still preserving a sufficient number of signal
events. A careful analysis of such events at the hadronic B-factories can
therefore reveal useful information enabling one to ascertain the
correctness of different theoretical claims regarding the decay of
the charm quark with a b-spectator. 

\vskip 1 true cm

We acknowledge helpful discussions with Tariq Aziz and Sheldon Stone.

\end{document}